\documentstyle
[12pt,dina4,epsf]{article}
\begin{document}
\parindent=0cm
\parskip=1.5mm

\def\bi{\begin{list}{$\bullet$}{\parsep=0.5\baselineskip
      \topsep=\parsep \itemsep=0pt}} \def\ei{\end{list}}
\def\phi{\varphi} \def\-{{\bf --}} \def\vm{v_{max}}
\newcommand{\eins}{\mathbf 1}
\begin{center}
  {\LARGE\bf The asymmetric exclusion model}
\end{center}
\begin{center}
  {\LARGE\bf with sequential update}
\end{center}
\vskip1.8cm \setcounter{footnote}{0}
\begin{center}
  {\Large N.\ Rajewsky\footnote{email: \tt nr@thp.uni-koeln.de}, A.\
    Schadschneider\footnote{email: \tt as@thp.uni-koeln.de} and M.\
    Schreckenberg\footnote{email: \tt schreck@rs1.uni-duisburg.de} }
\end{center}
\vskip1.3cm
\begin{center}
  $^{1,2}$\,Institut f\"ur Theoretische Physik\\ Universit\"at zu K\"oln\\
  D--50937 K\"oln, F.R.G.
\end{center}
\begin{center}
  $^{3}$ Theoretische Physik/FB 10\\
  Gerhard-Mercator-Universit\"at Duisburg\\
  D--47048 Duisburg, F.R.G.\\
\end{center}
\vskip1.3cm \vskip2cm {\large \bf Abstract}\hspace{.4cm} We present a
solution for the stationary state of an asymmetric exclusion model
with sequential update and open boundary conditions. We solve the
model exactly for random hopping in both directions by applying a
matrix-product formalism which was recently used to solve the model
with sublattice-parallel update \cite{hinri}. It is shown that the
matrix-algebra describing the sequential update and sublattice-parallel
 update are identical and can be mapped onto the
random sequential case treated by Derrida et al\cite{derrida93}.
\vfill \pagebreak The one-dimensional asymmetric exclusion model (AEM)
is one simple example of a reaction-diffusion model and has been used
to describe various problems in different fields of interest, like the
kinetics of bio polymerization {\cite {bio}} and traffic {\cite
  {traffic}}. Using recursion relations on the system size it was
solved 1992 {\cite {derrida92}} for the case of random sequential
update and open boundary conditions. Open boundaries here and in the
following mean that particles are injected at one end of a chain of
$L$ sites with probability $\alpha$ and removed at the other end with
probability $\beta$.  This model was then solved again by Derrida et
al 1993 {\cite {derrida93}} using a matrix product Ansatz (MPA),
inspired by the Matrix Product Groundstate for quantum spin chains
with groundstate energy zero {\cite {zitt}}, for the weights of the
stationary configurations. This Ansatz was most elegantly used to
obtain expressions for the density profile and higher correlations
consisting only of products of two matrices (belonging to the two
possible states of each site) and two vectors describing the influence
of the boundaries. Since then the MPA was extended to find also the
transient of the model {\cite {stinch}} and to recover solutions of
certain known integrable reaction-diffusion models {\cite {resolve}}.
All these models work with random sequential update.
\\ Hinrichsen{\cite {hinri}} solved the AEM model with
sublattice-parallel update
and could confirm earlier conjectures for the correlation functions
{\cite {schuetz2}}. It should be made clear that this update is
substantially different from the fully parallel update used for
example for modelling traffic flow. Nevertheless, to our knowledge
this is the first model with noncontinous time which has been solved
using the MPA. It is an open question for which classes of models the
MPA can be used successfully. \\
In this letter we study the AEM with {\it sequential} update. The
model then can be
defined as follows:\\
Consider sites located on a chain of length $L$. Each site $i$ ($0\le
i \le L$) can be occupied by a particle ($\tau_i=1$) or may be empty
($\tau_i=0$).  We start the update at the right end of the chain and
remove a particle at site $i=L$ with probability $\beta$. We then look
at the pair of sites $(i=L-1,i=L)$. If we find a particle at site
$L-1$ and no particle (called a hole) at site $L$ we move the particle
one site to the right with probability p. In the opposite case the
particle hops one site to the left with probability q. In the
remaining two cases nothing happens. We continue the update with the
pair $(i=L-2,i=L-1)$ and so forth until we reach the left end of the
chain. After the update of pair $(i=1,i=2)$ we inject a particle at
site $i=1$ with probability $\alpha$ if the site is empty.\\
Let us remark at this point that this model put on a ring (no
injection/removing of particles and periodic boundary conditions) has
a trivial stationary state  \cite{prep,diss} where correlations
are absent.\\
One could reverse
the order of the sequential update and go from the left to the right
through the chain (using the same rule as before for the pair-update).
These two models are connected by a particle-hole symmetry: injecting
particles can be seen as removing holes, and vice versa. Therefore it
is sufficient to study just one model.  The particle-hole symmetry
gives a first hint at the phase diagram: one would naively expect the
same phase diagram for both models. This implies a symmetry of the
phase diagram in $\alpha $ and $\beta $.\\
Following the MPA, we write
the stationary probability distribution
$P_0(\tau_1,\tau_2,...,\tau_L)$ as
\begin{equation}
  P_0(\tau_1,\tau_2,...,\tau_L)=Z_L^{-1}\langle W|\prod_{i=1}^L
  (\tau_i D + (1-\tau_i) E)|V\rangle
\end{equation}
or formally as
\begin{equation}
  |P_0\rangle \;=\;{Z_L}^{-1}\; \langle W| \, \left( \hspace{-1.5mm}
    \begin{array}{c} E \\[-1mm] D \end{array} \hspace{-1.5mm} \right)
  ^ {\otimes L} |V\rangle\,\,.
\end{equation}
The square matrices $E$ and $D$ can be infinite-dimensional. The
vectors $\langle W|$ and $|V\rangle$ act in the same vectorspace as
$E$ and $D$ and will, like $E$ and $D$, depend on $\alpha$ and
$\beta$.  The normalization constant is given by $Z_L=\langle
W|(D+E)^L |V\rangle$.  $|P_0\rangle$ represents the weights of all the
configurations in the stationary state.  This implies that
$|P_0\rangle$ is invariant under the action of the update-operator or
transfer matrix $T$:
\begin{equation}
  T \, |P_0\rangle = |P_0\rangle
\end{equation}
Let us now explicitly write down $T$. The boundary conditions may be
represented by operators ${\cal R}$ and ${\cal L}$ acting on site
$i=L$ and $i=1$, respectively:
\begin{equation}
{\cal R} \;=\; \left(
\begin{array}{cc}
  1 & \beta \\ 0 & 1-\beta
\end{array} \right) \,,
\hspace{10mm}
{\cal L} \;=\; \left(
\begin{array}{cc}
  1-\alpha & 0 \\ \alpha & 1
\end{array} \right)\,.
\end{equation}
The chosen basis for ${\cal R}$ and ${\cal L}$ is $(0,1)$. The
update-rule for any pair of sites $(i,i+1)$ can be written as
\begin{equation}
\label{Interaction}
{\cal T}_{i} \;=\; \left(
\begin{array}{cccc}
  1 & 0 & 0 & 0 \\ 0 & 1-q & p & 0 \\ 0 & q & 1-p & 0 \\ 0 & 0 & 0 & 1
\end{array} \right)\,.
\end{equation}
The basis is $(00,01,10,11)$ and we have formally
\begin{equation}
  T=L\cdot T_1\cdot ... \cdot T_{(L-1)}\cdot R
\end{equation}
with
\begin{eqnarray}
L &=& {\cal L}\otimes\eins\otimes ... \otimes\eins\,,\\
R &=& \eins\otimes ... \otimes\eins\otimes{\cal R}\,,\\
T_i &=& \eins\otimes\eins ...
\otimes{\cal T}_{i}\otimes\eins ... \otimes\eins\,,
\end{eqnarray}
where $\eins$ denotes the identity matrix.\\
\vfill\newpage
It is peculiar that one can use exactly the same mechanism which was
used to determine the stationary state of the model
with sublattice-parallel update {\cite {hinri}}
in order to fulfil equation (3):
\def\edmatrix{\Bigl(\hspace{-1.5mm}
    \begin{array}{c} E \\[-1mm] D
    \end{array}\hspace{-1.5mm} \Bigr)}
\def\abmatrix{\Bigl(\hspace{-1.5mm}
    \begin{array}{c} \hat{E} \\[-1mm] \hat{D}
    \end{array}\hspace{-1.5mm} \Bigr)}
\begin{equation}
\label{Equations}
{\cal T} \, \left[ \edmatrix \otimes \abmatrix \right] \;=\;
\abmatrix \otimes \edmatrix  \,,
\end{equation}
$$
\langle W| {\cal L} \abmatrix \;=\; \langle W| \edmatrix\,,
\hspace{15mm}
{\cal R} \edmatrix |V\rangle\;=\; \abmatrix|V\rangle \,,
$$
with some square matrices $\hat{E},\hat{D}$. This means that a 'defect'
is created in the beginning of an update at site $i=L$ and then transported
through the chain until it reaches the left end, where it disappears.\\
Equation (10) leads to the following bulk algebra:
\begin{eqnarray}
\label{BulkAlgebra}
[E,\hat{E}] =[D,\hat{D}] &=& 0 \quad ,\nonumber \\
(1-q)E\hat{D}+pD\hat{E} &=& \hat{E}D \quad ,\\
qE\hat{D}+(1-p)D\hat{E} &=& \hat{D}E \quad ,\nonumber
\end{eqnarray}
and the boundary conditions
\begin{equation}
\label{BoundaryConditions}
\begin{array}{c}
   \langle W| \hat{E} (1-\alpha) \;=\; \langle W| E\quad ,  \\[1mm]
   \langle W| (\alpha \hat{E}+\hat{D}) \;=\; \langle W| D \quad, \end{array}
\hspace{20mm}
\begin{array}{c}
   (1-\beta)D|V\rangle \;=\; \hat{D}|V\rangle\quad , \\[1mm]
   (E+\beta D)|V\rangle \;=\; \hat{E}|V\rangle \quad .\end{array}
\end{equation}
For $q=0$, $p=1$ we recover the algebra which
was solved in \cite{hinri}
for  the model with sublattice parallel-update, and we
can adopt the
two-dimensional representations of $E,D,\hat{E},\hat{D},\langle W|,|V\rangle$
of the algebra (7), even though $|P_0\rangle$ has a different structure.
One can easily show that the density profile of the sequential update
corresponds to the density of the even sites with sublattice-parallel update
and one gets essentially the same phase diagram.\\
One can check that for the case
\begin{equation}
(1-\alpha)(1-\beta)(1-q)=1-p
\end{equation}
one finds a one-dimensional solution of the complete algebra. This equation
defines the lines in the phase diagram on which the mean field solution
becomes exact. One can use these lines to calculate different currents
in the phase diagram \cite{prep,diss} and  we can exclude the case
$p=q=1$ in the following. \\
To solve the general algebra, we first note that
\begin{equation}
[E+D,\hat{E}+\hat{D}]=0
\end{equation}
holds for all values of $p,q$.
By demanding
\begin{eqnarray}
\hat{E}&=& E+\lambda \eins \,,\\
\hat{D} &=&D-\lambda \eins
\end{eqnarray}
(with some real number $\lambda$ and the identity matrix $\eins$) one can
reduce the whole algebra of seven equations to just three equations:
\begin{eqnarray}
pDE-qED &=& \lambda(1-q)E+\lambda(1-p)D\nonumber\,,\\
\alpha\langle W|E &=& \lambda(1-\alpha)\langle W|\,,\\
\beta D|V \rangle &=& \lambda |V \rangle \nonumber.
\end{eqnarray}
We define
\begin{eqnarray}
\widetilde{D}&:=&\lambda(1-p)D \,,\nonumber\\
\widetilde{E}&:=&\lambda(1-q)E \,, \\
\lambda^2&:=&\frac{1}{(1-q)(1-p)}\,, \nonumber
\end{eqnarray}
and rewrite eq.(14) as
\begin{eqnarray}
p\widetilde{D}\widetilde{E}-q\widetilde{E}\widetilde{D} &=&
\widetilde{E}+ \widetilde{D}\,, \nonumber\\
\alpha (1-p)\langle W| \widetilde{E}&=& (1-\alpha)\langle W|\,,\\
\beta (1-q)\widetilde{D}|V \rangle &=& |V \rangle.\nonumber
\end{eqnarray}
This is the algebra for the AEM with random sequential update but the
same local transfer matrix and the same boundary conditions as in our model.
It was solved by Derrida et al \cite{derrida93} with infinite-dimensional
matrices. Note that one has to rescale the vectors $\langle W|$ and
$|V\rangle$ of their solution with $\frac{1-\alpha}{1-p}$ and
$\frac{1}{1-q}$, respectively. \\
For the case $q=0$ we write down a slightly different
representation ($\lambda=1$):
\begin{equation}
{D} \;=\; \frac{1}{p}\left(
\begin{array}{cccccc}
\frac{p}{\beta} & a_1 & 0 & 0 & \cdot \\
0 & 1 & 1 & 0 & \cdot \\
0 & 0 & 1 & 1 & \cdot \\
0 & 0 & 0 & 1 & \cdot \\
\cdot & \cdot & \cdot & \cdot & \cdot \\
\end{array} \right)\,,
{E} \;=\; \frac{1}{p}\left(
\begin{array}{cccccc}
\frac{p(1-\alpha)}{\alpha} & 0 & 0 & 0 & \cdot \\
a_2 & 1-p & 0 & 0 & \cdot \\
0 & 1-p & 1-p & 0 & \cdot \\
0 & 0 & 1-p & 1-p & \cdot \\
\cdot & \cdot & \cdot & \cdot & \cdot \\
\end{array}
\right)\,,\end{equation}\begin{equation}
\langle W|=(1,0,0,0),\quad |V\rangle=\left( \hspace{-1.5mm}
\begin{array}{c} 1 \\
0\\ 0\\ 0\\ \end{array} \hspace{-1.5mm}  \right)\quad ,
\end{equation}
\begin{equation}
a_1a_2=\frac{p}{\alpha\beta}[(1-\alpha)\beta +\alpha -p]\,.
\end{equation}
The MPA makes it now straightforward to calculate the density profile, higher
correlations, the current and the phase diagram. We note at this point that
for the case $q=0$ the randomness in $p$ produces qualitatively
the same phase diagram as in the case of  random sequential dynamics with
$q=0$, $p=1$. Further results, also
concerning the comparison of the AEM with random sequential,
sublattice-parallel, sequential and fully parallel update will be
published elsewhere \cite{prep,diss}.\\

The present work shows that the sequential AEM with stochastic
hopping in both directions can be solved by means
of a mapping onto the known solution of the random sequential AEM.
This implies  also a generalization of the known solution of the AEM for
sublattice-parallel update, which had been  solved for deterministic hopping.
It is remarkable  that $pDE-qED=E+D$ is the  fundamental bulk algebra equation
for all three types of update discussed so far.
It is now possible to understand more about the nature and implications
of different types of update and randomness\cite{prep,diss}.\\
The solution of the sequential dynamics gives reason to hope that the
MPA is capable of solving the fully parallel case, which can be written as
a three-state model with the same sequential update as studied in
this work\cite{prep,diss}.\\
  \\
  \\
\noindent
{\bf Acknowledgements}\\[2mm]
This work has been performed within the research program of the
Sonderforschungsbereich 341 (K\"oln-Aachen-J\"ulich).
We would like to thank L.~Santen and B.~Strocka
for fruitful discussions. While writing the manuscript, we were informed
about investigations on the model with sublattice update and
random hopping performed by Peschel et al \cite{peschl}.

\end{document}